\documentclass[12pt]{article}
\usepackage[utf8]{inputenc}
\usepackage[left=1in,right=1in]{geometry}
\usepackage{authblk}
\usepackage[pdfstartview=FitH,pdfpagemode=None]{hyperref}
\usepackage{amsmath}
\usepackage{amssymb}
\usepackage{amsfonts}
\usepackage{graphicx}
\usepackage{braket}
\usepackage{cite}

\title{Black Holes and Marchenko-Pastur Distribution}
\author[1,2]{Wolfgang M\"uck}
\affil[1]{Dipartimento di Fisica ``Ettore Pancini'', Universit\`a degli Studi di Napoli Federico II \authorcr Via Cintia, 80126 Napoli, Italy}
\affil[2]{Istituto Nazionale di Fisica Nucleare, Sezione di Napoli \authorcr Via Cintia, 80126 Napoli, Italy}
\date{}

\begin{document}


\numberwithin{equation}{section}

\newcommand{\ie}{i.e.,\ }
\newcommand{\eg}{e.g.,\ }

\newcommand{\const}{\operatorname{const.}}

\newcommand{\sgn}{\operatorname{sgn}}

\newcommand{\rmd}{\,\mathrm{d}}

\newcommand{\Tr}{\operatorname{tr}}


\newcommand{\re}{\operatorname{Re}}
\newcommand{\im}{\operatorname{Im}}

\newcommand{\e}[1]{\operatorname{e}^{#1}}


\newcommand{\vev}[1]{\left\langle #1 \right\rangle}

\newcommand{\op}{\mathcal{O}}
\newcommand{\Liou}{\mathcal{L}}
\newcommand{\Hilb}{{\mathcal{H}}}

\newcommand{\Order}{\mathcal{O}}

\newcommand{\unit}[1]{\operatorname{#1}}

\newcommand{\BesselJ}[1]{\operatorname{J}_{#1}}
\newcommand{\BesselI}[1]{\operatorname{I}_{#1}}

\newcommand{\dbar}[1]{\bar{\bar{#1}}}

\newcommand{\Jacobi}[1]{P_{#1}^{(\alpha,\beta)}}
\newcommand{\hypF}[1]{\operatorname{{}_2F_1}\!\left(#1\right)}
\newcommand{\genhypF}[3]{\operatorname{{}_{#1}F_{#2}}\!\left(#3\right)}

\maketitle

\begin{abstract}
The universal eigenvalue distribution characterizing the Gram matrix of semiclassical ensembles of black hole microstates is recognized as the Marchenko-Pastur distribution, which plays a prominent role as the universal limit distribution in a large class of random matrix and vector models. It is proposed that this distribution also universally determines the energy spectral density of black holes, which allows to construct a Krylov space for the time evolution of typical black hole states and calculate their state complexity. It is checked that the state complexity growth at late times saturates Lloyd's bound. Some implications of the proposed spectral density for the generation of Hawking radiation and black hole evaporation are discussed. 
\end{abstract}
\section{Introduction}

In a recent series of papers \cite{Balasubramanian:2022gmo, Balasubramanian:2022lnw, Climent:2024trz}, the Gram matrix $\langle \Psi_i|\Psi_j \rangle$ of a generic large family of black hole microstates $F_\Omega = \{|\Psi_i\rangle \in \mathcal{H}: i=1,2,\ldots, \Omega\}$ was calculated for black holes in asymptotically AdS or Minkowski spacetimes, with or without electric charge and angular momenta, as well as for certain supersymmetric black holes. The microstates were constructed using thin matter shells in a heavy-shell limit and with the appropriate charges, similar to Wheeler's ``bag of gold'' configurations \cite{Wheeler:1964qna}. Specifically, the heavy-shell limit served the purpose of making the families as generic as possible and placing the microstates at infinite distance from each other in the semiclassical phase space. Despite of this, connected wormhole geometries with multiple boundaries generate non-trivial overlaps in the gravitational path integral that evaluates the moments of the Gram matrix.

In all cases discussed in \cite{Balasubramanian:2022gmo, Balasubramanian:2022lnw, Climent:2024trz}, the Gram matrix possesses a spectral density given by 
\begin{equation}
\label{bh:spec.dens}
	D(\lambda) = \delta(\lambda) \left(\Omega - \e{S} \right) I_{\{\Omega>\e{S}\}} 
	+ \frac{\e{S}}{2\pi \lambda} \sqrt{\left(\lambda-\lambda_-\right)\left(\lambda_+-\lambda\right)} I_{\{\lambda\in(\lambda_-, \lambda_+)\}}~,
\end{equation}
where 
\begin{equation}
\label{bh.entropy}
	S = \frac{A}{4G\hbar}
\end{equation}
is the Bekenstein-Hawking black hole entropy. $I_{\{E\}}$ denotes the event indicator function,\footnote{The event indicator function is defined by 
$$
	I_{\{event\}} = \begin{cases} 
				1 & \text{if $event$ is true,}\\
				0 & \text{if $event$ is false.}
			\end{cases}
$$
Alternatively, one may write it in terms of the Heavyside $\Theta$ function.}
and the limits of the continuum part of the spectrum are 
\begin{equation}
\label{lambda.pm}
	\lambda_\pm = \left(1 \pm \sqrt{\Omega \e{-S}} \right)^2~.
\end{equation}
The rank of the Gram matrix, \ie the number of non-zero eigenvalues, counts the independent microstates within $F_{\Omega}$ and turns out to be $d_\Omega = \min \left(\e{S}, \Omega\right)$. This result provides a very strong statistical interpretation of the Bekenstein-Hawking formula.

The distribution \eqref{bh:spec.dens} is, apart from the normalization, nothing but the Marchenko-Pastur (MP) distribution \cite{Marchenko:1967}, 
\begin{equation}
\label{comp:MP.distrib}
	f(\lambda) = \delta(\lambda) \left(1 -\frac1c \right) I_{\{c>1\}} 
	+ \frac{1}{2\pi c \lambda} \sqrt{\left(\lambda-\lambda_-\right)\left(\lambda_+-\lambda\right)} I_{\{\lambda\in(\lambda_-, \lambda_+)\}}~,
\end{equation}
with $\lambda_\pm = \left(1 \pm \sqrt{c} \right)^2$. More precisely, $D(\lambda)=\Omega f(\lambda)$ with $c=\Omega\e{-S}\in (0,\infty)$. 
The significance of the MP distribution stems from the MP theorem \cite{Marchenko:1967}, which can be stated as follows. Let $X$ be an $M \times N$ matrix with independent and identically distributed real or complex entries. Then, in the large $N$ limit with $M/N \to c \in (0,\infty)$ fixed, the asymptotic eigenvalue distribution of the $M\times M$ matrix $A =\frac1N X X^\dagger$ (also called the \emph{Wishart matrix}) is universal and given by the MP distribution. See \cite{Liu:2018hlr} for an example in the context of random matrix theory.
The MP distribution also appears under the name of \emph{free Poisson distribution} in Free Probability Theory \cite{Speicher:2009}. 
Since MP's original proof from 1967, many different proofs have appeared in the mathematical literature (see \cite{Haagerup:2003, Ledoux:2004, Yaskov:2016} and references therein). For physicists, a proof using Feynman diagrams \cite{Lu:2014jua} seems especially interesting. 

The MP theorem also finds an application in random vector models \cite{DeCock:1999, DeCock:2000}. Consider a sequence $\Phi=(\phi_1, \phi_2,\ldots, \phi_K)$ of randomly generated vectors in an $N$-dimensional Hilbert space $\mathcal{H}$, and let $\Gamma^\Phi$ be the Gram matrix of the sequence $\Phi$. Such sequences can be thought of as mimicking the dynamics of chaotic systems. The rank of $\Gamma^\Phi$ coincides with the number of independent vectors in $\Phi$. In the limit $N,K\to \infty$, with $K/N\to c$, the spectral distribution of $\Gamma^\Phi$ converges to the MP distribution. 

Bearing in mind the above statements, it is perhaps not surprising that one universally encounters the MP distribution when studying the quantum properties of black holes. Black holes are maximally chaotic thermal systems \cite{Hayden:2007cs, Sekino:2008he, Shenker:2013pqa, Maldacena:2015waa, Turiaci:2018zsj}, and essential aspects of chaos and black hole physics can be studied using random matrix models \cite{Mehta:2004, Balasubramanian:2014gla, Roberts:2016hpo, Cotler:2016fpe, Cotler:2017jue, Kar:2021nbm, Hacker:2023}.  

In this paper, we shall extend the results of \cite{Balasubramanian:2022gmo, Balasubramanian:2022lnw, Climent:2024trz} by proposing a linear relation between (binding) energy and the variable $\lambda$, such that the MP distribution with parameter $c\approx 1$ universally determines the semiclassical energy spectral density of black holes. This places a semiclassical black hole ensemble at the threshold between densities with purely continuous spectra and spectra consisting of a continuum and a discrete energy level. Having a spectral density allows to construct a Krylov space and a one-dimensional hopping model, which captures in precise detail the time evolution of a certain initial quantum state. In particular, it allows for an unambiguous\footnote{For Krylov (operator) complexity, there is some ambiguity related to the choice of an operator inner product \cite{Kar:2021nbm, Muck:2022xfc}. For states in a Hilbert space, we assume that the inner product is unique, so that state complexity depends only on the initial state.} definition of state complexity \cite{Balasubramanian:2022tpr}. The relation between state complexity and Nielson's notion of circuit complexity has been discussed in \cite{Chattopadhyay:2023fob, Aguilar-Gutierrez:2023nyk}. The late-time behaviour of state complexity is generic for typical states in a chaotic system. 
We will show that the growth of state complexity in our model saturates Lloyd's bound \cite{Lloyd:2000} at late times. 
The proposed model also provides for a simple mechanism of the emission of Hawking radiation. 
We will discuss some implications and check that the mean emitted energy is compatible with the Hawking temperature. 

The remainder of the paper is structured as follows. In section~\ref{comp}, the general framework of Krylov space and state complexity is reviewed. In section~\ref{BH}, this framework is applied to the MP distribution. Our proposal for the relation between the MP distribution and a universal energy spectral density for black hole ensembles is presented and discussed in section~\ref{BH.toy}. Finally, section~\ref{conc} contains the conclusions.

\section{Krylov state complexity}
\label{comp}

Krylov state complexity \cite{Balasubramanian:2022tpr}, sometimes called spread complexity, is a measure for how fast a state spreads out in Hilbert space as it evolves in time. It is a generalization of operator, or Krylov, complexity \cite{Parker:2018yvk}, which measures the spread of an observable in the space of operators and was originally proposed as an alternative to out-of-time-ordered correlators (OTOCs) to measure chaos. 
In this section, the mathematical framework of state complexity will be reviewed following the very nice expositions in \cite{Balasubramanian:2022tpr, Alishahiha:2022anw, Erdmenger:2023wjg}. A similar analysis restricted to Krylov (operator) complexity was presented in \cite{Kar:2021nbm, Muck:2022xfc}.

\subsection{Krylov space}

In a quantum system described by a Hilbert space $\mathcal{H}$ and a time-independent Hamiltonian $H$, the unitary time evolution of a state $\ket{\psi(t)} \in \mathcal{H}$ in the Schrödinger picture is given by
\begin{equation}
	\label{comp:Schroedinger}
	\ket{\psi(t)} = \e{-itH}\ket{\psi(0)}~.
\end{equation}
Formally, the exponential can be expanded into an infinite series involving the states $\ket{\psi_n}=H^n \ket{\psi(0)}$, $n=0,1,2,\ldots$. These states span a Hilbert space $\mathcal{K} \subset \mathcal{H}$, which is the minimal subspace of $\mathcal{H}$ containing the entire trajectory of $\ket{\psi(t)}$. Let $d_{\mathcal{H}}$ and $d_{\mathcal{K}}$ denote the dimensions of $\mathcal{H}$ and $\mathcal{K}$, respectively. Clearly, one has $d_{\mathcal{K}}\leq d_{\mathcal{H}}$. $\mathcal{K}$ is called the Krylov subspace. 

However, the $\ket{\psi_n}$ defined above do not form a nice basis of $\mathcal{K}$; they need neither be orthogonal to nor independent of each other. The Gram-Schmidt procedure that produces an ordered, orthogonal basis of $\mathcal{K}$, which is called the Krylov basis, is known as the Lanczos algorithm \cite{Lanczos:1950, Viswanath-Mueller}. The essence of the Lanczos algorithm is captured by the three-term recurrence relation\footnote{This is the \emph{monic} version of the recurrence relation giving rise to monic orthogonal polynomials. In the literature, one more often finds the version that produces \emph{normalized} polynomials and states \cite{Balasubramanian:2022tpr, Erdmenger:2023wjg}. It is, however, easy to keep track of the normalization.} 
\begin{equation}
\label{comp:Kn.recurrence}
	H\ket{K_n} = \ket{K_{n+1}} + a_n \ket{K_n} +\Delta_n \ket{K_{n-1}}~,
\end{equation}
with the initial state $\ket{K_0}=\ket{\psi(0)}$ and the coefficients $a_n$ and $\Delta_n$ given by
\begin{equation}
\label{comp:a.Delta}
	a_n = \frac{\braket{K_n|H|K_n}}{h_n}~,\quad \Delta_n = \frac{h_n}{h_{n-1}}~~(n>0)~, 
	\quad h_n = \braket{K_n|K_n}~,\quad \Delta_0=0~. 
\end{equation}
If $d_{\mathcal{K}}<\infty$, the Lanczos algorithm terminates with $h_{d_{\mathcal{K}}}=0$.

Given \eqref{comp:Kn.recurrence}, it is now possible to explicitly write 
\begin{equation}
\label{comp:K.P}
	\ket{K_n} = P_n(H) \ket{K_0}~,
\end{equation} 
where $P_n$ is a monic orthogonal polynomial of degree $n$ satisfying the same recurrence relation,
\begin{equation}
	\label{comp:Pn.recurrence}
	EP_n(E) = P_{n+1}(E) + a_n P_n(E) +\Delta_n P_{n-1}(E)~, 
\end{equation}
and the orthogonality relations 
\begin{equation}
\label{comp:Pn.ortho}
	\int \rmd \mu(E) P_n(E) P_m(E) = h_n \delta_{mn}~.
\end{equation}
The measure $\mu(E)$ appearing in \eqref{comp:Pn.ortho} is implied by  
\begin{equation}
\label{comp:measure}
	\int \rmd \mu(E) f(E) = \braket{K_0| f(H) |K_0}
\end{equation}   
and coincides with the heuristic measure
\begin{equation}
\label{comp:measure.equiv}
	\frac{\rmd \mu}{\rmd E} = \sum\limits_n \delta(E-E_n) \left|\braket{E_n| K_0}\right|^2~,
\end{equation}   
where the sum is over all eigenstates $\ket{E_n}$ of the Hamiltonian.

\subsection{Hopping model}

The fact that the Krylov basis is an \emph{ordered} basis is plays a crucial role in what follows. 
Define the expansion coefficients of $\ket{\psi(t)}$ in the Krylov basis,
\begin{equation}
\label{comp:phi.n.def}
	\phi_n(t) = \frac{i^n}{\sqrt{h_n}} \braket{K_n|\psi(t)}~. 
\end{equation}
They obviously satisfy\footnote{We do not include evolution in imaginary time, which was considered in \cite{Erdmenger:2023wjg}.}
\begin{equation}
\label{comp:phi.conserved}
	\sum\limits_{n} |\phi_n(t)|^2 = \sum\limits_{n} \frac{\braket{\psi(t)|K_n}\braket{K_n|\psi(t)}}{h_n}
	= \braket{\psi(t)|\psi(t)} = 1 
\end{equation}
and may be rewritten, using \eqref{comp:measure}, as
\begin{equation}
\label{comp:phi.n.rewrite}
	\phi_n(t) = \frac{i^n}{\sqrt{h_n}} \braket{K_0|P_n(H) \e{-iHt} |K_0} = i^n \int\rmd\mu(E) \frac{P_n(E)}{\sqrt{h_n}} \e{-iEt}~.
\end{equation}

Then, the Schrödinger equation for $\ket{\psi(t)}$ translates into the following hopping equation on the one-dimensional chain of wavefunctions \cite{Viswanath-Mueller},
\begin{equation}
\label{comp:phi.eq}
	\partial_t \phi_n(t) = -\sqrt{\Delta_{n+1}} \phi_{n+1}(t) -i a_n \phi_{n}(t) + \sqrt{\Delta_{n}} \phi_{n-1}(t)~.
\end{equation} 
The initial condition of this hopping problem is $\phi_n(0) =\delta_{n,0}$. 

It is evident from \eqref{comp:phi.eq} that the function $\phi_0(t)$, which is called the \emph{survival amplitude}, contains all the information about the entire chain of wavefunctions. An important task is to calculate it from the knowledge of the Lanczos coefficients. In the classic work \cite{Viswanath-Mueller}, this problem is approached by introducing the Laplace transform of $\phi_n$, 
\begin{equation}
\label{comp:cn.def}
	c_n(z) = \int\limits_{0}^\infty \phi_n(t) \e{-zt}\qquad (\Re z >0)~,
\end{equation} 
in terms of which \eqref{comp:phi.eq} becomes
\begin{equation}
\label{comp:c.eq}
	z c_n(z) = -\sqrt{\Delta_{n+1}} c_{n+1}(z) -i a_n c_{n}(z) + \sqrt{\Delta_{n}} c_{n-1}(z) + \delta_{n,0}~.
\end{equation} 

In a slightly more modern language, the $c_n(z)$ are closely related to the \emph{functions of the second kind} \cite{vanAsche:1990, Grinshpun:2013}, 
\begin{equation}
\label{comp:Qn.def}
	Q_n(z) = \int \rmd\mu(E) \frac{P_n(E)}{z-E}~,\qquad z\in \mathbb{C}\backslash\text{supp}(\mu(E))~.  
\end{equation}
More precisely, inserting \eqref{comp:phi.n.rewrite} into \eqref{comp:cn.def} yields 
\begin{equation}
\label{comp:cQ.rel}
	c_n(z) = \frac{i^{n+1}}{\sqrt{h_n}} Q_n(iz)~,
\end{equation}
which also provides the analytic continuation of \eqref{comp:cn.def} to $\Re z<0$. 
The $Q_n(z)$ satisfy, apart from the case $n=0$, the same recurrence relation as the $P_n(E)$, 
\begin{equation}
\label{comp:Q.recurrence}
	zQ_n(z) = Q_{n+1}(z) + a_n Q_n(z) +\Delta_n Q_{n-1}(z) +\delta_{n,0}~.
\end{equation}

The function $Q_0$ is the \emph{resolvent},
\begin{equation}
\label{comp:Q0.resolv}
	Q_0(z) = \Braket{K_0|\frac{1}{z-H}|K_0}~.
\end{equation}
It encodes the spectrum and the measure via 
\begin{equation}
\label{comp:measure.Q}
	\frac{\rmd\mu}{\rmd E} = \frac1{2\pi i} \lim\limits_{\epsilon\to 0^+} \left[ Q_0(E-i\epsilon) - Q_0(E+i\epsilon) \right]~. 
\end{equation}

A continuous fraction representation of the resolvent can be found from \eqref{comp:Q.recurrence} as follows. Defining 
\begin{equation}
\label{comp:Rn.def}
	R_n(z) = \frac{Q_n(z)}{Q_{n-1}(z)}\qquad (n>0)~,
\end{equation}
\eqref{comp:Q.recurrence} can be rewritten as 
\begin{equation}
\label{comp:Q0.cont.frac}
	Q_0(z) = \frac1{z-a_0-R_1(z)}~,\qquad  R_{n}(z) = \frac{\Delta_n}{z-a_n-R_{n+1}(z)}~.
\end{equation}
This yields
\begin{equation}
\label{comp:Q0.cont.frac2}
	Q_0(z) = \frac1{z-a_0-\frac{\Delta_1}{z-a_1-\frac{\Delta_2}{z-a_2-\cdots}}}~.
\end{equation}

\subsection{State complexity}
\label{comp:state.comp}

State complexity is defined as the expectation value of the position $n$ along the chain of wavefunctions,
\begin{equation}
\label{comp:C.hopping}
	C(t) = \sum\limits_{n=0}^\infty n |\phi_n(t)|^2~. 
\end{equation}
Because the index $n$ also counts the basis vectors in Krylov space, this definition suggests that state complexity should be understood as an effective dimension, not as a distance. In fact, the distance axioms are in general not satisfied by state complexity \cite{Aguilar-Gutierrez:2023nyk}.   

Following a similar treatment for Krylov complexity \cite{Muck:2022xfc}, we will elaborate on this formula with the purpose of providing an expression that is more useful for practical calculations. 
Consider the Laplace transform of $C(t)$,
\begin{equation}
	\label{comp:C.Laplace}
	\tilde{C}(z) = \int\limits_0^\infty \rmd t\, C(t) \e{-zt} \qquad (\Re z >0)~.
\end{equation}
After substituting \eqref{comp:C.hopping} and \eqref{comp:phi.n.rewrite} and integrating, this becomes
\begin{equation}
	\label{comp:C.L2}
	\tilde{C}(z) = \sum\limits_{n=1}^\infty n \int\rmd \mu(E) \int \rmd \mu(E') \frac{P_n(E)P_n(E')}{h_n[z+i(E-E')]}~.
\end{equation}
One may recognize the function of the second kind \eqref{comp:Qn.def}, so that
\begin{equation}
	\label{comp:C.L3}
	\tilde{C}(z) = -i \sum\limits_{n=1}^\infty n \int\rmd \mu(E) \frac{P_n(E)Q_n(-iz+E)}{h_n}~.
\end{equation}
The factor $n$ makes this sum difficult to compute, but the recurrence relations \eqref{comp:Pn.recurrence} and \eqref{comp:Q.recurrence} can be used to obtain a handier expression. The trick is to multiply by $z$, write $-iz =(-iz+E)-E$, and then use \eqref{comp:Pn.recurrence} and \eqref{comp:Q.recurrence}. After shifting the summation index in some of the terms, this results in
\begin{equation}
	\label{comp:C.L4}
	z \tilde{C}(z) = \sum\limits_{n=0}^\infty \int\rmd \mu(E) \frac{P_{n+1}(E)Q_n(-iz+E)-P_{n}(E)Q_{n+1}(-iz+E)}{h_n}~.
\end{equation}
Multiplying once more by $z$ and applying the same procedure, one obtains
\begin{align}
	\label{comp:C.L5}
	z^2 \tilde{C}(z) &= i \sum\limits_{n=0}^\infty \int\rmd \mu(E) \left\{\frac{2(\Delta_n-\Delta_{n+1})}{h_n}P_n(E)Q_n(-iz+E) \right. \\
	\notag &\quad
	\left. + \frac{a_n-a_{n+1}}{h_n}\left[P_n(E)Q_{n+1}(-iz+E) + P_{n+1}(E)Q_n(-iz+E)\right] \right\}~.
\end{align}

The analogue of \eqref{comp:C.L5} in real time, \ie its inverse Laplace transform, was called the \emph{Ehrenfest theorem} for complexity in \cite{Erdmenger:2023wjg}. An instructive way to derive it is as follows. The complexity \eqref{comp:C.hopping} may also be written as
\begin{equation}
\label{comp:comp.op}
	C(t) = \braket{\psi(0)|\mathcal{K}(t)|\psi(0)}~,\qquad 
	\mathcal{K}(t) = \e{iHt} \sum\limits_{n=0}^\infty n \frac{\ket{K_n}\bra{K_n}}{h_n} \e{-iHt}~,
\end{equation}
where $\mathcal{K}(t)$ is the complexity operator in the Heisenberg picture. Obviously, 
\begin{equation}
\label{comp:Ehrenfest}
	\partial^2_t \mathcal{K}(t) = - [H,[H,\mathcal{K}(t)]]~.
\end{equation}
The right hand side of \eqref{comp:Ehrenfest} can be manipulated using the recurrence relations, and taking the expectation value gives the Ehrenfest theorem.

To conclude this section, we will make a few general statements based on the properties of orthogonal polynomials \cite{Chihara:1978, Koornwinder:2013fcq}. These statements will address, in particular, the generic late-time behaviour of state complexity. 

If $\mathcal{K}$ is finite-dimensional, then the spectrum is discrete and finite, and $d_{\mathcal{K}}$ is an upper bound of the complexity $C(t)$. At late times, $C(t)$ may approach a constant value, $C(t)\to C_\infty < d_{\mathcal{K}}$, known as the plateau \cite{Barbon:2019wsy, Rabinovici:2020ryf}, or it may oscillate, even maximally between zero and $d_\mathcal{K}$. Therefore, the presence of a late-time plateau in physical systems, in contrast to unlimited growth, is a simple consequence of the fact that any physical system ultimately has a finite number of degrees of freedom. If $d_\mathcal{K}$ is huge, it will typically take an exponentially long time to reach the plateau. A late-time plateau with more or less pronounced oscillations also appears to occur in cases with unbounded discrete spectra \cite{Muck:2022xfc}.

Systems with a very large number of degrees of freedom can be effectively described in a large-$N$ (thermodynamic) limit for time scales that are short compared to the time needed to approach the plateau. The continuum approximation is a useful tool \cite{Barbon:2019wsy, Alishahiha:2022anw, Muck:2022xfc, Erdmenger:2023wjg} to study such systems. In the large-$N$ limit, the spectrum becomes continuous and may be either bounded or unbounded. The following general statements can be made \cite{Chihara:1978, Koornwinder:2013fcq}. The spectrum is bounded if and only if both sets of Lanczos coefficients, $\Delta_n$ and $a_n$, are bounded. If, for $n\to \infty$, these coefficients approach limits, $\Delta_n\to \bar{\Delta}$ and $a_n\to \bar{a}$, then the spectrum is bounded with at most countably many points outside the interval $[\bar{a}-2\sqrt{\bar{\Delta}}, \bar{a}+2\sqrt{\bar{\Delta}}]$,  and the limits of this interval are limit points of the spectrum. The MP distribution is an example of such a spectrum and will be discussed in section~\ref{BH}.

\subsection{Linear transformation of energy}
\label{comp:lin.map}

This subsection addresses the impact of the linear map
\begin{equation}
\label{comp:E.lin.map}
	H' = \alpha H + \beta~, \qquad \alpha \neq 0~,
\end{equation} 
with real constants $\alpha$ and $\beta$, on the quantities characterizing Krylov space and state complexity. 
We emphasize that the case of negative $\alpha$ is not excluded, which is possible as long as the spectrum is bounded. 
The linear map \eqref{comp:E.lin.map} induces the following transformations:
\begin{align}
\label{comp:trafo.P}
	P'_n(E') &= \alpha^n P_n(E)~, \qquad \ket{K'_n} = \alpha^n \ket{K_n}~,\\
\label{comp:trafo.a.Delta}
	a'_n &= \alpha a_n + \beta~, \qquad \Delta'_n = \alpha^2 \Delta_n~,\\
\label{comp:trafo.h}
	h'_n &= \alpha^{2n} h_n~,\\
\label{comp:trafo.Q}
	Q'_n(\alpha z + \beta) &= \alpha^{n-1} Q_n(z)~.
\end{align}  
Clearly, the \emph{normalized} Krylov basis states remain unchanged. However, the time-dependent wave functions $\phi_n$ transform by
\begin{equation}
\label{comp:trans.phi}
	\phi'_n(t) = \e{-i\beta t} 
	\begin{cases}
		\phi_n(\alpha t) &\text{for $\alpha >0$,}\\
		\phi_n^\ast(|\alpha|t) \quad &\text{for $\alpha <0$.}
	\end{cases}
\end{equation}
The distinction between the two cases is necessary, because the time domain is always $t\geq0$.
It follows that the complexity transforms according to
\begin{equation}
\label{comp:trans.C}
	C'(t) = C(|\alpha|t)~, \qquad \tilde{C}'(z) = \frac1{|\alpha|} \tilde{C}\left(\frac{z}{|\alpha|}\right)~.
\end{equation}

\section{Krylov space of the MP distribution}
\label{BH}

\subsection{MP distribution from Lanczos algorithm}

Consider the following simple set of Lanczos coefficients,  
\begin{equation}
\label{BH:Lanczos}
	\Delta_n = c~,\qquad a_n = 1 + c \quad (n>0)~, \qquad a_0=1~,
\end{equation} 
where $c\geq 0$ is a free parameter.  
In this subsection, it will be shown that these Lanczos coeffients are associated with the MP distribution discussed in the introduction. 
For uniformity of notation, $\lambda$ will be used as the formal energy variable instead of $E$.  

Consider the resolvent $Q_0(z)$, which can be calculated with the help of \eqref{comp:Q0.cont.frac}. More precisely, with \eqref{BH:Lanczos} one has $R_n(z)=R(z)$ for $n>0$, which satisfies
\begin{equation}
\label{BH:R.eq}
	R(z) = \frac{c}{z-(1+c)-R(z)}~.
\end{equation}
This yields
\begin{equation}
\label{BH:R.sol}
	R(z) = \frac12 \left[z-1-c \pm \sqrt{(z-1-c)^2 - 4c} \right]~,
\end{equation}
which, in turn, implies 
\begin{align}
\notag 	
	Q_0(z) &= \frac1{z-1-R(z)} \\
\label{BH:Q0.sol}
	&= \frac{1}{2c z} \left[z-1+c \pm \sqrt{\left(z-1-c\right)^2 - 4c} \right]~.
\end{align}
The sign is determined by the condition $Q_0(z)\to 0$ for $|z|\to \infty$. 
The spectral density that follows from \eqref{comp:measure.Q} is the MP distribution \eqref{comp:MP.distrib}. 

For completeness, we report the following moments,
\begin{align}
\label{BH:E1}
	\braket{\lambda} &= \int \rmd \mu\, \lambda = 1~,\\
\label{BH:E2}
	\braket{\lambda^2} &= \int \rmd \mu\, \lambda^2 = 1+c~.
\end{align}

\subsection{Late-time state complexity}
\label{BH:late.time.complexity}

Using the general framework of section~\ref{comp}, the Lanczos coefficients \eqref{BH:Lanczos} give rise to a Krylov basis and a chain of wavefunctions, to which one can assign a state complexity \eqref{comp:C.hopping}. This is the state complexity associated with the MP distribution. 

In this subsection, we shall calculate the late-time linear growth coefficient of this state complexity. Recall that $\lambda$ formally parameterizes energy, but one should keep in mind the possibility of a linear transformation, as discussed in subsection~\ref{comp:lin.map}. The late-time behaviour of $C(t)$ is encoded in the small-$z$ expansion of the Laplace transform $\tilde{C}(z)$. In particular, a leading order behaviour $\tilde{C}(z)\sim \frac{C_0}{z^2}$ implies the late-time complexity $C(t) \sim C_0 t$. This is relevant in the present case.

The starting point is the general formula \eqref{comp:C.L5}. With the Lanczos coefficients \eqref{BH:Lanczos}, the sum reduces to the $n=0$ summand,
\begin{equation}
	\label{BH:Cz.1}
	z^2 \tilde{C}(z) = i \int\rmd \mu(\lambda) \left\{ -2c Q_0(-iz+\lambda) -c \left[ Q_1(-iz+\lambda) + P_1(\lambda)Q_0(-iz+\lambda)\right] \right\}~.
\end{equation}
Using $Q_1(z) = (z-1)Q_0(z) -1$ from \eqref{comp:Q.recurrence} and $P_1(\lambda)=\lambda-1$ from \eqref{comp:Pn.recurrence}, \eqref{BH:Cz.1} becomes
\begin{equation}
	\label{BH:Cz.2}
	z^2 \tilde{C}(z) = i c \left[ 1 + \int\rmd \mu(\lambda) (iz-2\lambda) Q_0(-iz+\lambda)\right]~.
\end{equation}
Furthermore, substituting $Q_0$ using its definition \eqref{comp:Qn.def} yields
\begin{equation}
	\label{BH:Cz.3}
	z^2 \tilde{C}(z) = -i c \int\rmd \mu(\lambda) \int \rmd\mu(\lambda') \frac{\lambda+\lambda'}{-iz +\lambda-\lambda'}~.
\end{equation}

To continue, one must take care of the delta function part of the measure. For this purpose, formally write \eqref{comp:MP.distrib} as  
\begin{equation}
\label{BH:spec.dens.short}
	\frac{\rmd \mu}{\rmd \lambda} =  (1-A) \frac{\rmd \tilde{\mu}}{\rmd \lambda} + A \delta(\lambda)~, \qquad 
	A= \left(1-\frac1{c} \right) I_{\{c>1\}}~,
\end{equation}
where the measure $\tilde{\mu}$ is normalized on the continuum. 
With this abbreviation, \eqref{BH:Cz.3} becomes 
\begin{equation}
	\label{BH:Cz.4}
	z^2 \tilde{C}(z) = -i c \left[ A(1-A) \int\rmd \tilde{\mu}(\lambda) \frac{2iz\lambda}{z^2+\lambda^2} 
	+ (1-A)^2 \int \rmd \tilde{\mu}(\lambda) \int \rmd\tilde{\mu}(\lambda') \frac{\lambda+\lambda'}{-iz +\lambda-\lambda'} \right]~.
\end{equation}

At this point, the limit $z\to0$ can be taken in order to extract the coefficient $C_0$. The measure $\tilde{\mu}$ has support for $\lambda>0$, so that the first term in the bracket vanishes in this limit. The fraction in the second term reduces to a principal value and a delta function, but the integral with the contribution from the principle value vanishes by symmetry. The surviving term stems from the delta function and reads 
\begin{align}
\notag
	C_0=\lim\limits_{z\to 0} z^2 \tilde{C}(z) &= 2\pi c (1-A)^2 \int\limits_{\lambda_-}^{\lambda_+} \rmd \lambda\left( \frac{\rmd\tilde{\mu}}{\rmd \lambda} \right)^2 \lambda\\
\notag 
	&= \frac{1}{2\pi c} \int\limits_{\lambda_-}^{\lambda_+} \rmd \lambda \frac{(\lambda-\lambda_-)(\lambda_+-\lambda)}{\lambda} \\
\notag 
	&= \frac{1}{4\pi c} \left( \lambda_+^2 -\lambda_-^2 -2\lambda_+\lambda_- \ln\frac{\lambda_+}{\lambda_-}\right)\\ 
\label{BH:Cz.5}
	&= \frac{2}{\pi} \left[ c^{\frac12} + c^{-\frac12} 
	- \frac12\left(c^{\frac12} - c^{-\frac12} \right)^2 \ln\left|\frac{c^{\frac12}+1}{c^{\frac12} -1}\right| \right]~,
\end{align}
where \eqref{lambda.pm} has been used in the last step.

\begin{figure}[th]
	\begin{center}
		\includegraphics[width=0.75\textwidth]{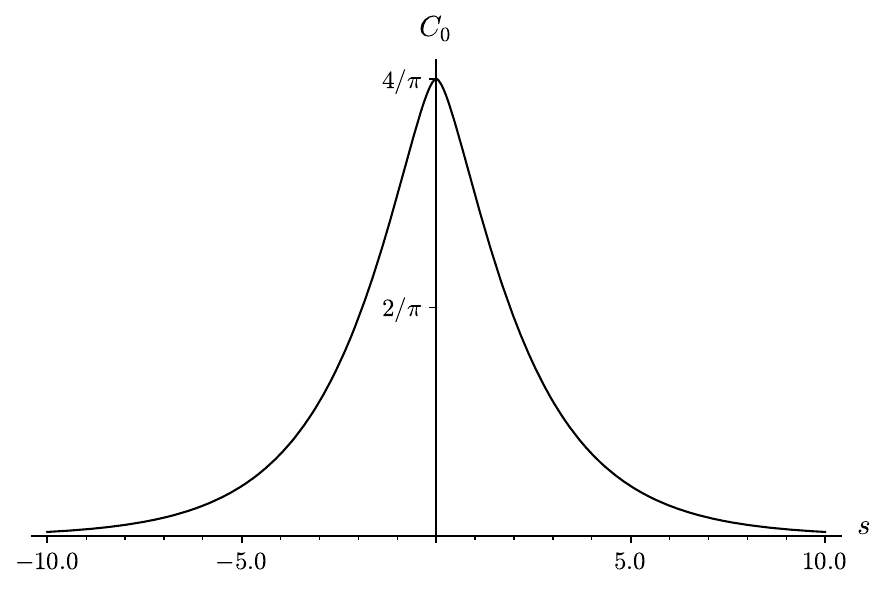}
	\end{center}
	\caption{Plot of $C_0$ as a function of $s$.
		\label{BH:c0.plot}}
\end{figure}

A small change of variable puts this into a more suggestive form.
Defining $s$ by   
\begin{equation}
\label{BH:s.def}
	c = \e{s}~,
\end{equation}
\eqref{BH:Cz.5} becomes
\begin{equation}
\label{BH:c0.s}
	C_0 = \frac{4}{\pi} \left[ \cosh\left(\frac{s}{2}\right) + \sinh^2\left(\frac{s}{2}\right) \ln \left| \tanh \frac{s}{4} \right| \right]~.
\end{equation}
This is an even function of $s$, sharply peaked at $s=0$ with $C_{0,\text{max}}=\frac4{\pi}$, and asymptotically decays as $C_0\sim \mathcal{O}(\e{-\frac12|s|})$ for $|s|\to \infty$. A graph is shown in Figure~\ref{BH:c0.plot}. 

\section{Black hole ensemble}
\label{BH.toy}

As discussed in the introduction, the MP distribution appears in the context of black hole physics as the eigenvalue distribution of the Gram matrix of generic families of black hole microstates. In this section, we will extrapolate this result and propose that the MP distribution with parameter $c\approx 1$, in fact, captures also the distribution of (gravitational) binding energy of a black hole, which is seen as a many-body bound state. The state complexity associated with the MP distribution according to the general framework of state evolution in Krylov space is shown to saturate Lloyd's bound. We shall also discuss some implications of this proposal for black hole evaporation.

\subsection{MP distribution and black hole spectrum}

Recall that the MP distribution appears as the universal limit distribution of the eigenvalues of the Gram matrix for an ensemble of size $\Omega$ of microstates that are macroscopically indistinguishable from a black hole with given mass $M$ and possibly other charges, which determine its macroscopic entropy $S$ \cite{Balasubramanian:2022gmo, Balasubramanian:2022lnw, Climent:2024trz}. The absence or presence of zero eigenvalues signals whether the microstates in the ensemble are independent of each other or not. 
These two cases exactly occur for $\ln\Omega<S$ ($c<1$) and $\ln\Omega>S$ ($c>1$), respectively. 
This immediately leads to the conclusion that an ensemble truly representing a black hole in or close to equilibrium should have the MP distribution with parameter $c\approx 1$ as its limit distribution. We will call such an ensemble a black hole ensemble. 
In terms of the parameter $s$ defined in \eqref{BH:s.def}, a black hole ensemble is characterized by 
\begin{equation}
\label{BH:s.interpret}
	s = \ln \Omega - S\approx 0~.
\end{equation} 

\begin{figure}[th]
	\begin{center}
		\includegraphics[width=0.6\textwidth]{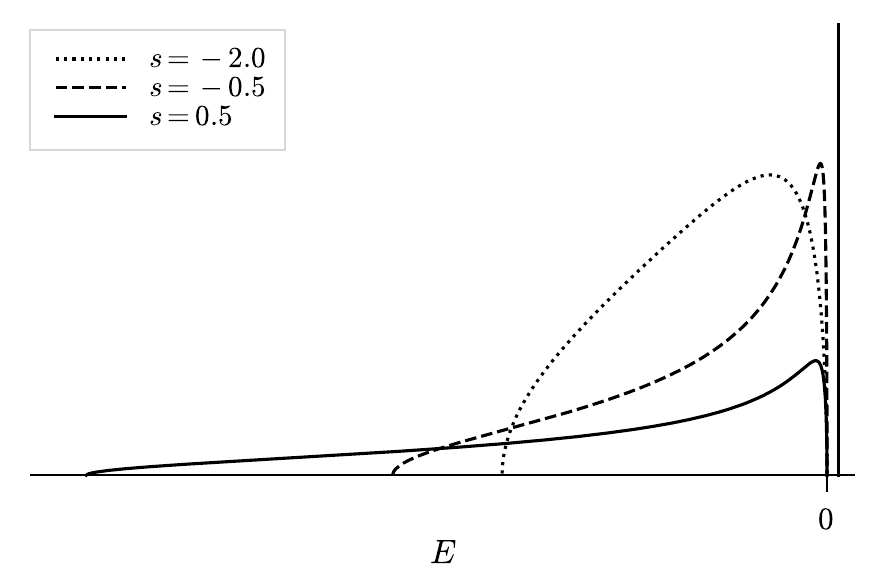}
	\end{center}
	\caption{Illustration of spectral densities $\frac{\rmd{\mu}}{\rmd E}$ for various values of $s$. The solid vertical line represents the delta function for $s=0.5$.
		\label{BH:densplot}}
\end{figure}

For the time being, let us ignore this specific value and consider $s$ as a free parameter. 
Consider some large physical system and assume that it has a spectrum of energies given by
\begin{equation}
\label{BH:E.model}
	E = -M\frac{\lambda-\lambda_-}{1-\lambda_-}~,  
\end{equation}
where $\lambda$ is distributed according to the MP distribution. This particular map from $\lambda$ to $E$ has been chosen such that the continuum part of the spectrum describing a (gravitationally) bound state has $E=0$ as its upper bound, and the total binding energy is\footnote{In black hole physics, the fact that the binding energy is \emph{minus} the total mass $M$ is supported by the Brown-York quasi-local energy \cite{Brown:1992br}, or by the teleparallel approach to gravity \cite{Lucas:2009nq}. The energy contained in the gravitational field outside a Schwarzschild black hole with ADM mass $M$ is $-M$. The bare energy contribution from inside the black hole radius is $2M$.} 
\begin{equation}
\label{BH:vev.E}
	\vev{E} = -M~.
\end{equation} 
The variance of energy is 
\begin{equation}
\label{BH:variance.E}
	\vev{E^2}-\vev{E}^2 =  \frac{M^2}{(2-\e{s/2})^2}~.
\end{equation} 
Furthermore, the delta function part of the spectrum, which appears for $s>0$, is located at 
\begin{equation}
\label{BH:E.delta}
	E = \omega_s = \frac{M\lambda_-}{1-\lambda_-}>0~.
\end{equation}
Obviously, the relations above require $\lambda_-<1$, which restricts $s$ by $s<2\ln 2$. This restriction is actually unessential, because, as will be discussed in the next subsection, $s>0$ represents unstable systems, and the range of parameters needed to describe the physical process of black hole evaporation is limited to small, positive values of $s$. 
It can be observed that, in the limit $s\to-\infty$ ($c\to0$), the distribution becomes a semi-circle distribution on $E\in(-2M,0)$, whereas in the opposite physical limit, $s\to 0^-$ ($c\to 1^-$), one has $E\in(-4M,0)$ with a sharp peak at $E=0$.
Some examples are illustrated in figure~\ref{BH:densplot}.

Given the spectrum, it is now possible to associate a Krylov basis with it according to the general framework described in section~\ref{comp}. This also provides the Krylov state complexity, which can be interpreted as the complexity of a typical state in the ensemble. In particular, the late-time complexity growth is
$C(t) \approx  \dot{C} t$, where the growth rate $\dot{C}$ is found from \eqref{comp:trans.C}, \eqref{BH:c0.s} and \eqref{BH:E.model} as 
\begin{equation}
\label{BH:CE}
	\dot{C} = \frac{4M}{\pi\e{s/2}\left(2-\e{s/2}\right)} \left[ \cosh\left(\frac{s}{2}\right) + \sinh^2\left(\frac{s}{2}\right) \ln \left| \tanh \frac{s}{4} \right| \right]~.
\end{equation}
\begin{figure}[th]
	\begin{center}
		\includegraphics[width=0.75\textwidth]{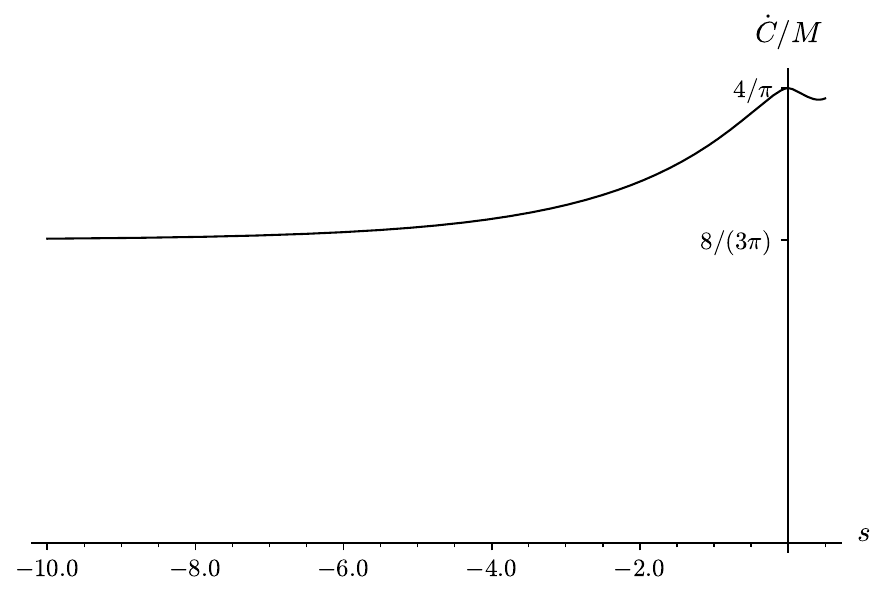}
	\end{center}
	\caption{Plot of $\dot{C}/M$ as a function of $s$. 
		\label{BH:CE.plot}}
\end{figure}
A plot of $\dot{C}$ is shown in figure~\ref{BH:CE.plot}. Clearly, the rescaling from $\lambda$ to $E$ breaks the symmetry $s\leftrightarrow -s$ that was present in \eqref{BH:c0.s}. At the same time, it changes the asymptotic behaviour for $s\to -\infty$, which is now
\begin{equation}
\label{BH:CE.asympt}
	\dot{C} \to \frac{8M}{3\pi} \qquad \text{for $s\to -\infty$.}
\end{equation}
However, the value at $s=0$ remains unchanged, which is also the maximum for $s\leq 0$~, 
\begin{equation}
\label{BH:CE.max}
	\dot{C}_{\text{max}} = \frac{4M}{\pi}~.	
\end{equation}
This value saturates Lloyd's bound \cite{Lloyd:2000}, in agreement with the interpretation of $s=0$ as the value appropriate for a black hole ensemble. We mention that, for positive $s$, $\dot{C}$ has a local minimum barely visible at the end of the plot and diverges for $s=2\ln 2$. However, as mentioned above, this parameter region is outside the physical range of interest.

\subsection{Black hole evaporation}

The entropy $S$ of a black hole with a certain mass $M$ and possibly other charges is the maximum entropy any physical system with the same mass and charges can possess. Therefore, interpreting the parameter $s$ in \eqref{BH:s.interpret} as a measure of the entropy of an ensemble compared to the maximum attainable, one must conclude that systems with $s>0$ are physically unstable. In the context of black hole ensembles, which lie at the threshold between stable and unstable systems, this instability leads to Hawking radiation and black hole evaporation.

Because $s$ is a semiclassical parameter, it is natural to think that it is subject to quantum fluctuations. 
Consider now what would happen if, by means of vacuum fluctuations, $s$ reached some small, positive value. Recall that for $s>0$, the spectrum has a delta function contribution, and the gap above the continuum is 
\begin{equation}
\label{BH:gap}
	\omega_s = \frac{M\lambda_-}{1-\lambda_-} \approx \frac{Ms^2}{4}~.
\end{equation} 
In the last relation $\lambda_-=(\e{s/2}-1)^2$ has been expanded for small $s$. 
This discrete part of the spectrum represents a quantum of Hawking radiation ready to be emitted. 
The model does not provide the probability for the emission of Hawking radiation with exactly this energy (\ie at this particular value of $s$), but it reproduces the correct scaling relations with $M$. 
Identifying the weight of the delta function in the distribution [see \eqref{BH:spec.dens.short}] with the number of Hawking particles emitted per unit of time, the luminosity, \ie the energy radiated by the black hole in a unit of time, would be given by
\begin{equation}
\label{BH:E.emitted}
	L = A \omega_s =\left(1-\e{-s}\right) \omega_s \approx \frac{Ms^3}{4}~.
\end{equation}
This should be compared to the luminosity of a black hole. Let us omit numerical factors for simplicity. For a Schwarzschild black hole \cite{Page:1976df}, $L\sim1/M^2$ in Planck units. Together with \eqref{BH:E.emitted}, this implies the scaling relations 
\begin{equation}
\label{BH:s.M.scaling}
	s\sim \frac1{M}~,\qquad \omega_s\sim \frac1{M}~.
\end{equation} 
Clearly, the scaling of $\omega_s$, \ie the mean energy of a single emitted particle, agress with the Hawking temperature, as expected.

A qualitative picture of black hole evaporation can now be sketched as follows. 
Semiclassically, a black hole ensemble in thermal equilibrium has a limit distribution characterized by $s\to 0^-$. This value is subject to quantum fluctuations. As long a $s$ stays negative, nothing special happens, because the spectrum only contains bound states. However, when $s$ becomes positive, the distribution undergoes a phase transition in which states with $E>0$ appear. These may be emitted as Hawking radiation. After emission, the remainder of the black hole is again described by a distribution with a continuous spectrum of bound states (but with a slightly lower mass). It will quickly rearrange by thermal equilibration to a limit distribution with $s\to 0^-$, so that the whole process can repeat itself.

\section{Conclusions}
\label{conc}

In summary, the universal eigenvalue distribution of the Gram matrix for semiclassical ensembles of typical black hole microstates found in \cite{Balasubramanian:2022gmo, Balasubramanian:2022lnw, Climent:2024trz} has been recognized as the MP distribution, which is a well known limit distribution for random matrix and vector models. 
We have extended this result by proposing a spectral density for ensembles representing black holes of mass $M$. The proposed spectral density is based on the MP distribution for $c=0$, \ie the distribution at the threshold between a purely continuous spectrum and a continuum plus a discrete eigenvalue. 
This has allowed us to construct a Krylov basis and calculate the state complexity of a black hole. State complexity is found to grow linearly at late times saturating Lloyd's bound, which is an important check. 
An interesting observation is that the variance of energy in the proposed ensemble scales as $M^2$, which is different from a canonical ensemble, such as the thermofield double state, in which case it would scale as $M$. However, if one assumes that there exists a universal energy spectral density of black holes, up to a simple rescaling of energy relative to the black hole mass, then the variance must scale as $M^2$. This is also consistent with the scaling of the late-time complexity growth rate discussed in subsection~\ref{comp:lin.map}. 

The fact that the proposed black hole ensemble lies at the threshold between two qualitatively different spectra provides for an interesting possibility to generate and emit Hawking radiation. Indeed, it is natural to think that quantum fluctuations will occasionally push the system beyond threshold, leading to Hawking radiation and black hole decay. The scaling of the mean energy (black hole temperature) and luminosity in terms of $M$ have been checked for the proposed model and turn out to be consistent.
This mechanism of black hole evaporation is reminiscent of another proposal on the nature of black holes known as the quantum $N$-portrait \cite{Dvali:2011aa,Dvali:2012en}, although we have not been concerned with the microscopic details behind the quantum fluctuations. It would certainly be interesting to develop specific microscopic models that exhibit the proposed limit spectral density and allow for a deeper understanding of the evolution of black holes.

\section*{Acknowledgements}
I thank Yi Yang for interesting discussions. 
Partial support by the INFN under the research initiative STEFI is gratefully acknowledged.

\bibliographystyle{utphys}
\bibliography{complex}

\providecommand{\href}[2]{#2}\begingroup\raggedright\begin{thebibliography}{10}

\bibitem{Balasubramanian:2022gmo}
V.~Balasubramanian, A.~Lawrence, J.~M. Magan, and M.~Sasieta, ``{Microscopic
  Origin of the Entropy of Black Holes in General Relativity},''
  \href{http://dx.doi.org/10.1103/PhysRevX.14.011024}{{\em Phys. Rev. X}
  {\bfseries 14} no.~1, (2024) 011024},
  \href{http://arxiv.org/abs/2212.02447}{{\ttfamily arXiv:2212.02447
  [hep-th]}}.

\bibitem{Balasubramanian:2022lnw}
V.~Balasubramanian, A.~Lawrence, J.~M. Magan, and M.~Sasieta, ``{Microscopic
  origin of the entropy of astrophysical black holes},''
  \href{http://arxiv.org/abs/2212.08623}{{\ttfamily arXiv:2212.08623
  [hep-th]}}.

\bibitem{Climent:2024trz}
A.~Climent, R.~Emparan, J.~M. Magan, M.~Sasieta, and A.~Vilar~L\'opez,
  ``{Universal Construction of Black Hole Microstates},''
  \href{http://arxiv.org/abs/2401.08775}{{\ttfamily arXiv:2401.08775
  [hep-th]}}.

\bibitem{Wheeler:1964qna}
J.~A. Wheeler, ``{Geometrodynamics and the issue of final state},'' in {\em
  {Les Houches Summer School of Theoretical Physics}: {Relativity, Groups and
  Topology}}, pp.~317--522.
\newblock 1964.

\bibitem{Marchenko:1967}
V.~Marchenko and L.~Pastur, ``{Distribution of eigenvalues for some sets of
  random matrices},''
  \href{http://dx.doi.org/10.1070/SM1967v001n04ABEH001994}{{\em Math. USSR
  Sbornik} {\bfseries 1} (1967) 457}.

\bibitem{Liu:2018hlr}
J.~Liu, ``{Spectral form factors and late time quantum chaos},''
  \href{http://dx.doi.org/10.1103/PhysRevD.98.086026}{{\em Phys. Rev. D}
  {\bfseries 98} no.~8, (2018) 086026},
  \href{http://arxiv.org/abs/1806.05316}{{\ttfamily arXiv:1806.05316
  [hep-th]}}.

\bibitem{Speicher:2009}
R.~Speicher,
  \href{http://dx.doi.org/10.1093/oxfordhb/9780198744191.013.22}{``{Free
  probability theory},''} in {\em {The Oxford Handbook of Random Matrix
  Theory}}, G.~Akemann, J.~Baik, and P.~Di~Francesco, eds.
\newblock Oxford University Press, 2015.
\newblock \href{http://arxiv.org/abs/0911.0087}{{\ttfamily arXiv:0911.0087
  [math.PR]}}.
\newblock \url{https://doi.org/10.1093/oxfordhb/9780198744191.013.22}.

\bibitem{Haagerup:2003}
U.~Haagerup and S.~Thorbjørnsen, ``{Random matrices with complex Gaussian
  entries},''
  \href{http://dx.doi.org/https://doi.org/10.1016/S0723-0869(03)80036-1}{{\em
  Expositiones Mathematicae} {\bfseries 21} no.~4, (2003) 293--337}.
  \url{https://www.sciencedirect.com/science/article/pii/S0723086903800361}.

\bibitem{Ledoux:2004}
M.~Ledoux, ``{Differential Operators and Spectral Distributions of Invariant
  Ensembles from the Classical Orthogonal Polynomials. The Continuous Case},''
  \href{http://dx.doi.org/10.1214/EJP.v9-191}{{\em Electronic Journal of
  Probability} {\bfseries 9} (2004) 177 -- 208}.
  \url{https://doi.org/10.1214/EJP.v9-191}.

\bibitem{Yaskov:2016}
P.~Yaskov, ``{A short proof of the Marchenko–Pastur theorem},''
  \href{http://dx.doi.org/10.1016/j.crma.2015.12.008}{{\em Comptes Rendus
  Mathematique} {\bfseries 354} no.~3, (Mar., 2016) 319–322}.
  \url{http://dx.doi.org/10.1016/j.crma.2015.12.008}.

\bibitem{Lu:2014jua}
X.~Lu and H.~Murayama, ``{Universal Asymptotic Eigenvalue Distribution of Large
  $N$ Random Matrices --- A Direct Diagrammatic Proof to Marchenko-Pastur Law
  ---},'' \href{http://arxiv.org/abs/1410.3503}{{\ttfamily arXiv:1410.3503
  [hep-th]}}.

\bibitem{DeCock:1999}
M.~De~Cock, M.~Fannes, and P.~Spincemaille, ``{On quantum dynamics and
  statistics of vectors},''
  \href{http://dx.doi.org/10.1088/0305-4470/32/37/306}{{\em Journal of Physics
  A: Mathematical and General} {\bfseries 32} no.~37, (Sep, 1999) 6547}.
  \url{https://dx.doi.org/10.1088/0305-4470/32/37/306}.

\bibitem{DeCock:2000}
M.~De~Cock, M.~Fannes, and P.~Spincemaille, ``{Quantum dynamics and Gram's
  matrix},'' \href{http://dx.doi.org/10.1209/epl/i2000-00163-6}{{\em Europhys.
  Lett.} {\bfseries 49} no.~4, (2000) 403--409}.
  \url{https://doi.org/10.1209/epl/i2000-00163-6}.

\bibitem{Hayden:2007cs}
P.~Hayden and J.~Preskill, ``{Black holes as mirrors: Quantum information in
  random subsystems},''
  \href{http://dx.doi.org/10.1088/1126-6708/2007/09/120}{{\em JHEP} {\bfseries
  09} (2007) 120}, \href{http://arxiv.org/abs/0708.4025}{{\ttfamily
  arXiv:0708.4025 [hep-th]}}.

\bibitem{Sekino:2008he}
Y.~Sekino and L.~Susskind, ``{Fast Scramblers},''
  \href{http://dx.doi.org/10.1088/1126-6708/2008/10/065}{{\em JHEP} {\bfseries
  10} (2008) 065}, \href{http://arxiv.org/abs/0808.2096}{{\ttfamily
  arXiv:0808.2096 [hep-th]}}.

\bibitem{Shenker:2013pqa}
S.~H. Shenker and D.~Stanford, ``{Black holes and the butterfly effect},''
  \href{http://dx.doi.org/10.1007/JHEP03(2014)067}{{\em JHEP} {\bfseries 03}
  (2014) 067}, \href{http://arxiv.org/abs/1306.0622}{{\ttfamily arXiv:1306.0622
  [hep-th]}}.

\bibitem{Maldacena:2015waa}
J.~Maldacena, S.~H. Shenker, and D.~Stanford, ``{A bound on chaos},''
  \href{http://dx.doi.org/10.1007/JHEP08(2016)106}{{\em JHEP} {\bfseries 08}
  (2016) 106}, \href{http://arxiv.org/abs/1503.01409}{{\ttfamily
  arXiv:1503.01409 [hep-th]}}.

\bibitem{Turiaci:2018zsj}
G.~J. Turiaci, {\em {Black Holes and Chaos}}.
\newblock PhD thesis, Princeton U., 7, 2018.

\bibitem{Mehta:2004}
M.~L. Mehta, {\em {Random Matrices}}, vol.~142 of {\em {Pure and Applied
  Mathematics}}.
\newblock Elsevier, 3~ed., 2004.

\bibitem{Balasubramanian:2014gla}
V.~Balasubramanian, M.~Berkooz, S.~F. Ross, and J.~Simon, ``{Black Holes,
  Entanglement and Random Matrices},''
  \href{http://dx.doi.org/10.1088/0264-9381/31/18/185009}{{\em Class. Quant.
  Grav.} {\bfseries 31} (2014) 185009},
  \href{http://arxiv.org/abs/1404.6198}{{\ttfamily arXiv:1404.6198 [hep-th]}}.

\bibitem{Roberts:2016hpo}
D.~A. Roberts and B.~Yoshida, ``{Chaos and complexity by design},''
  \href{http://dx.doi.org/10.1007/JHEP04(2017)121}{{\em JHEP} {\bfseries 04}
  (2017) 121}, \href{http://arxiv.org/abs/1610.04903}{{\ttfamily
  arXiv:1610.04903 [quant-ph]}}.

\bibitem{Cotler:2016fpe}
J.~S. Cotler, G.~Gur-Ari, M.~Hanada, J.~Polchinski, P.~Saad, S.~H. Shenker,
  D.~Stanford, A.~Streicher, and M.~Tezuka, ``{Black Holes and Random
  Matrices},'' \href{http://dx.doi.org/10.1007/JHEP05(2017)118}{{\em JHEP}
  {\bfseries 05} (2017) 118}, \href{http://arxiv.org/abs/1611.04650}{{\ttfamily
  arXiv:1611.04650 [hep-th]}}. [Erratum: JHEP 09, 002 (2018)].

\bibitem{Cotler:2017jue}
J.~Cotler, N.~Hunter-Jones, J.~Liu, and B.~Yoshida, ``{Chaos, Complexity, and
  Random Matrices},'' \href{http://dx.doi.org/10.1007/JHEP11(2017)048}{{\em
  JHEP} {\bfseries 11} (2017) 048},
  \href{http://arxiv.org/abs/1706.05400}{{\ttfamily arXiv:1706.05400
  [hep-th]}}.

\bibitem{Kar:2021nbm}
A.~Kar, L.~Lamprou, M.~Rozali, and J.~Sully, ``{Random matrix theory for
  complexity growth and black hole interiors},''
  \href{http://dx.doi.org/10.1007/JHEP01(2022)016}{{\em JHEP} {\bfseries 01}
  (2022) 016}, \href{http://arxiv.org/abs/2106.02046}{{\ttfamily
  arXiv:2106.02046 [hep-th]}}.

\bibitem{Hacker:2023}
P.~Hacker, ``{Complexity, chaos and black hole microstates},'' 2023.
\newblock (Ph.D. thesis, Vrije Universiteit Brussel).

\bibitem{Muck:2022xfc}
W.~M\"uck and Y.~Yang, ``{Krylov complexity and orthogonal polynomials},''
  \href{http://dx.doi.org/10.1016/j.nuclphysb.2022.115948}{{\em Nucl. Phys. B}
  {\bfseries 984} (2022) 115948},
  \href{http://arxiv.org/abs/2205.12815}{{\ttfamily arXiv:2205.12815
  [hep-th]}}.

\bibitem{Balasubramanian:2022tpr}
V.~Balasubramanian, P.~Caputa, J.~M. Magan, and Q.~Wu, ``{Quantum chaos and the
  complexity of spread of states},''
  \href{http://dx.doi.org/10.1103/PhysRevD.106.046007}{{\em Phys. Rev. D}
  {\bfseries 106} no.~4, (2022) 046007},
  \href{http://arxiv.org/abs/2202.06957}{{\ttfamily arXiv:2202.06957
  [hep-th]}}.

\bibitem{Chattopadhyay:2023fob}
A.~Chattopadhyay, A.~Mitra, and H.~J.~R. van Zyl, ``{Spread complexity as
  classical dilaton solutions},''
  \href{http://dx.doi.org/10.1103/PhysRevD.108.025013}{{\em Phys. Rev. D}
  {\bfseries 108} no.~2, (2023) 025013},
  \href{http://arxiv.org/abs/2302.10489}{{\ttfamily arXiv:2302.10489
  [hep-th]}}.

\bibitem{Aguilar-Gutierrez:2023nyk}
S.~E. Aguilar-Gutierrez and A.~Rolph, ``{Krylov complexity is not a measure of
  distance between states or operators},''
  \href{http://arxiv.org/abs/2311.04093}{{\ttfamily arXiv:2311.04093
  [hep-th]}}.

\bibitem{Lloyd:2000}
S.~Lloyd, ``{Ultimate physical limits to computation},''
  \href{http://dx.doi.org/10.1038/35023282}{{\em {Nature}} {\bfseries 406}
  (2000) 1047--1054}, \href{http://arxiv.org/abs/quant-ph/9908043}{{\ttfamily
  quant-ph/9908043}}.

\bibitem{Parker:2018yvk}
D.~E. Parker, X.~Cao, A.~Avdoshkin, T.~Scaffidi, and E.~Altman, ``{A Universal
  Operator Growth Hypothesis},''
  \href{http://dx.doi.org/10.1103/PhysRevX.9.041017}{{\em Phys. Rev. X}
  {\bfseries 9} no.~4, (2019) 041017},
  \href{http://arxiv.org/abs/1812.08657}{{\ttfamily arXiv:1812.08657
  [cond-mat.stat-mech]}}.

\bibitem{Alishahiha:2022anw}
M.~Alishahiha and S.~Banerjee, ``{A universal approach to Krylov state and
  operator complexities},''
  \href{http://dx.doi.org/10.21468/SciPostPhys.15.3.080}{{\em SciPost Phys.}
  {\bfseries 15} no.~3, (2023) 080},
  \href{http://arxiv.org/abs/2212.10583}{{\ttfamily arXiv:2212.10583
  [hep-th]}}.

\bibitem{Erdmenger:2023wjg}
J.~Erdmenger, S.-K. Jian, and Z.-Y. Xian, ``{Universal chaotic dynamics from
  Krylov space},'' \href{http://dx.doi.org/10.1007/JHEP08(2023)176}{{\em JHEP}
  {\bfseries 08} (2023) 176}, \href{http://arxiv.org/abs/2303.12151}{{\ttfamily
  arXiv:2303.12151 [hep-th]}}.

\bibitem{Lanczos:1950}
C.~Lanczos, ``{An Iteration Method for the Solution of the Eigenvalue Problem
  of Linear Differential and Integral Operators},''
  \href{http://dx.doi.org/10.6028%2Fjres.045.026}{{\em Journal of Research of
  the Nation al Bureau of Standards} {\bfseries 45} no.~4, (1950) 255--282}.

\bibitem{Viswanath-Mueller}
V.~Viswanath and G.~M\"uller, {\em {The Recursion Method}}, vol.~m 23 of {\em
  Lecture Notes in Physics}.
\newblock Springer-Verlag, 1994.

\bibitem{vanAsche:1990}
W.~van Asche, ``{Orthogonal polynomials, associated polynomials and functions
  of the second kind},'' {\em Journal of Computational and Applied Mathematics}
  {\bfseries 37} (1991) 237--249.

\bibitem{Grinshpun:2013}
E.~Grinshpun and Z.~Grinshpun, ``{On Functions of the Second Kind in Orthogonal
  Polynomial Theory},'' \href{http://dx.doi.org/10.1007/s40315-012-0006-7}{{\em
  Comput. Methods Funct. Theory} {\bfseries 13} (2013) 65--74}.

\bibitem{Chihara:1978}
T.~Chihara, {\em {An Introduction to Orthogonal Polynomials}}.
\newblock Gordon and Breach, 1978.
\newblock reprinted, Dover (2011).

\bibitem{Koornwinder:2013fcq}
T.~H. Koornwinder,
  \href{http://dx.doi.org/10.1007/978-3-7091-1616-6_6}{``{Orthogonal
  Polynomials},''} in {\em {LHCPhenoNet School}: {Integration, Summation and
  Special Functions in Quantum Field Theory}}, pp.~145--170.
\newblock 2013.

\bibitem{Barbon:2019wsy}
J.~L.~F. Barb\'on, E.~Rabinovici, R.~Shir, and R.~Sinha, ``{On The Evolution Of
  Operator Complexity Beyond Scrambling},''
  \href{http://dx.doi.org/10.1007/JHEP10(2019)264}{{\em JHEP} {\bfseries 10}
  (2019) 264}, \href{http://arxiv.org/abs/1907.05393}{{\ttfamily
  arXiv:1907.05393 [hep-th]}}.

\bibitem{Rabinovici:2020ryf}
E.~Rabinovici, A.~S\'anchez-Garrido, R.~Shir, and J.~Sonner, ``{Operator
  complexity: a journey to the edge of Krylov space},''
  \href{http://dx.doi.org/10.1007/JHEP06(2021)062}{{\em JHEP} {\bfseries 06}
  (2021) 062}, \href{http://arxiv.org/abs/2009.01862}{{\ttfamily
  arXiv:2009.01862 [hep-th]}}.

\bibitem{Brown:1992br}
J.~D. Brown and J.~W. York, Jr., ``{Quasilocal energy and conserved charges
  derived from the gravitational action},''
  \href{http://dx.doi.org/10.1103/PhysRevD.47.1407}{{\em Phys. Rev. D}
  {\bfseries 47} (1993) 1407--1419},
  \href{http://arxiv.org/abs/gr-qc/9209012}{{\ttfamily arXiv:gr-qc/9209012}}.

\bibitem{Lucas:2009nq}
T.~G. Lucas, Y.~N. Obukhov, and J.~G. Pereira, ``{Regularizing role of
  teleparallelism},'' \href{http://dx.doi.org/10.1103/PhysRevD.80.064043}{{\em
  Phys. Rev. D} {\bfseries 80} (2009) 064043},
  \href{http://arxiv.org/abs/0909.2418}{{\ttfamily arXiv:0909.2418 [gr-qc]}}.

\bibitem{Page:1976df}
D.~N. Page, ``{Particle Emission Rates from a Black Hole: Massless Particles
  from an Uncharged, Nonrotating Hole},''
\href{http://dx.doi.org/10.1103/PhysRevD.13.198}{{\em Phys. Rev.} {\bfseries
  D13} (1976) 198--206}.

\bibitem{Dvali:2011aa}
G.~Dvali and C.~Gomez, ``{Black Hole's Quantum N-Portrait},''
  \href{http://dx.doi.org/10.1002/prop.201300001}{{\em Fortsch. Phys.}
  {\bfseries 61} (2013) 742--767},
\href{http://arxiv.org/abs/1112.3359}{{\ttfamily arXiv:1112.3359 [hep-th]}}.

\bibitem{Dvali:2012en}
G.~Dvali and C.~Gomez, ``{Black Holes as Critical Point of Quantum Phase
  Transition},'' \href{http://dx.doi.org/10.1140/epjc/s10052-014-2752-3}{{\em
  Eur. Phys. J.} {\bfseries C74} (2014) 2752},
\href{http://arxiv.org/abs/1207.4059}{{\ttfamily arXiv:1207.4059 [hep-th]}}.

\end{thebibliography}\endgroup

\end{document}